\documentclass{article}

\PassOptionsToPackage{numbers, compress}{natbib}

\usepackage[final]{neurips_2022}

\usepackage[utf8]{inputenc} 
\usepackage[T1]{fontenc}    
\usepackage{hyperref}       
\usepackage{url}            
\usepackage{booktabs}       
\usepackage{amsfonts}       
\usepackage{nicefrac}       
\usepackage{microtype}      
\usepackage{xcolor}         
\usepackage{graphicx}
\usepackage{wrapfig}

\usepackage{enumitem}
\usepackage{subcaption}

\usepackage{comment}

\title{The Reasonable Effectiveness of \\ Diverse Evaluation Data}

\author{
  Lora Aroyo\\
  Google Research\\
  New York, US \\
  \texttt{l.m.aroyo@gmail.com} \\
   \And
   Mark Díaz \\
   Google Research\\
   New York, US \\
   \texttt{markdiaz@google.com} \\
   \And
   Christopher Homan \\
   Google Research\\
   New York, US \\
   \texttt{homanc@google.com} \\
   \AND
   Vinodkumar Prabhakaran \\
   Google Research\\
   San Francisco, US \\
   \texttt{vinodkpg@google.com} \\
   \And
   Alex Taylor \\
   Google Research\\
   London, UK \\
   \texttt{alxtyl@google.com} \\
   \And
   Ding Wang \\
   Google Research\\
   Singapore, Singapore \\
   \texttt{drdw@google.com} \\
}

\begin{document}
\maketitle

\begin{abstract}
In this paper, we present findings from an semi-experimental exploration of rater diversity and its influence on safety annotations of conversations generated by humans talking to a generative AI-chat bot. We find significant differences in judgments produced by raters from different geographic regions and annotation platforms, and correlate these perspectives with demographic sub-groups. Our work helps define best practices in model development-- specifically human evaluation of generative models-- on the backdrop of growing work on sociotechnical AI evaluations.

\end{abstract}

\section{Introduction}
\label{intro}
In their 2009 paper ``The Unreasonable Effectiveness of Data'' \cite{halevy2009unreasonable}, Alon Halevy, Peter Norvig, and Fernando Perreira urge ML researchers to ``follow the data and see where it leads.'' Since then, we have at an unprecedented scale amassed data for the purpose of machine learning. Yet, data quality---including the quality of human-collected data---has been left behind. 

While there is a substantial body of work focusing on the reliability of human raters when performing evaluations, there are few \cite{diaz2018addressing, luo2020detecting, goyal2022your,prabhakaran2021releasing} if any studies investigating how the characteristics of rater pools impact ratings. That is, we know little about annotators’ individual characteristics (such as nationality, gender, education, race) and how they might influence the way they label data. This matters because, in seeking to build fair and responsible AI systems, we should anticipate potential biases that may emerge as a result of differences across user populations, and evaluation data should represent a variety of populations in order to better reflect viewpoints among real-world stakeholders and build diversity aware ground truth datasets\cite{aroyo2015truth}. 

Responding to the \emph{Data-Centric AI} call to study impacts of data on AI systems \cite{Sambasivan_et_al}, we present findings from a semi-experimental exploration of rater diversity and its influence on safety annotations of chat bot conversations. We report results from a large-scale rater diversity study performed on a sample of 990 conversations generated by humans conversing with a generative AI-chatbot. We collected safety labels from a pool of 96 raters (recruited from two rating platforms covering a range of socio-economic subgroups) using 24 safety questions. Rather than requesting the typical number of ratings (a single rating or three-to-five ratings per conversation), we collected 40 ratings per safety question and repeated the experiment after 4 months. We analyzed:

\begin{itemize}[topsep=0pt,itemsep=0pt]
    \item the variance, disagreement and reliability within and across different demographic slices. 
    \item the majority and minority perspectives on the data items.
    \item the consistency of raters across replications of the experiment. 
\end{itemize}

The contribution and implications of our research are threefold. First, we propose a methodological approach for assessing the influences of rater diversity on data labels. Second, we demonstrate that raters' backgrounds do have an impact on the labels they provide. Third, we illustrate that there are discrepancies in rater consistency across two replications. As such, we contribute to two topics in this workshop: (1) repeatability of human evaluations and (2) methods for assessing the quality and the reliability of human evaluations. 

\section{Experimental Design for Diverse Human Evaluation}
\label{experiment}
The overall research question guiding our study was: Do rater demographics help us improve reliability of annotations and ultimately provide guidance for collecting representative data for AI development and evaluation?

\paragraph{Data:}
\label{data}
The input data for this study was sampled from a 8K conversations dataset (composing 48K turns in total) generated by humans interacting with a generative AI-chatbot. In our study, the sample of 990 conversations was annotated for 24 safety characteristics (plus one general question about conversation understandability) by 40 unique raters per conversation, per safety characteristic. This annotation experiment was repeated with a different pool of 96 raters (including 19 overlapping raters in both replications) 4 months later. The total number of annotations gathered is close to 2 million (990K in each repetition).  

\paragraph{Task:}
\label{task}
The safety evaluation task used a UI designed for the safety characteristics as defined in Appendix. Each rater was asked to rate the final chatbot response in each conversation as \emph{Unsafe}, \emph{Safe} or \emph{Unsure}. These answers were given for each of the 24 safety questions, which were organized in three groups (Q2-Q4 below), plus one question (Q1) checking the overall comprehensibility of the conversation.
\begin{itemize}[topsep=0pt,itemsep=0pt]
    \item Q1: Evaluate the whole conversation
    \item Q2: Evaluate the last bot utterance for “Harmful Content”
    \item Q3: Evaluate the last bot utterance for “Content with Unfair Bias”
    \item Q4: Evaluate the last bot utterance for “Misinformation and Political References”
\end{itemize}

\paragraph{Data Collection:}
\label{repetition}
We collected the ratings in two phases with an interval of four months. In both phases, we recruited 96 unique raters from two rater pools. All raters performed the task independently and used the same annotation template. All raters were asked to complete an optional demographic survey (e.g. gender, ethnicity, education level, age group, and native language). All questions in the demographic survey gave raters the option to select "Prefer not to answer". We also collected data about the average annotation time per conversation and the total time each rater spent annotating.  

\paragraph{Raters:}
\label{raters}
In Phase 1, the breakdown of raters was: 71 in Pool 1 (42 India, 29 US), 25 in Pool 2 (12 India, 13 US). For Phase 2, the break down was: 72 in Pool 1 (40 India, 32 US), 24 in Pool 2 (12 India, 12 US). 19 of the raters participated in both phases (5 from Pool 1, 14 from Pool 2; the 5 raters from Pool 1 are all in US, 6 out of the 14 from Pool 2 are in India and 8 in US; 9 identify as female and 10 as male). In this paper, we report results from the Phase 2, however we compare the two phases to measure consistency for the raters who participated in both. 

\section{Results}
\label{results}

We present four key high-level observations from this study that contribute to our understanding of reliability of human evaluations, and its relation to diversity among raters.

\begin{figure}[h!]
    \centering
    \includegraphics[width=0.43\textwidth]{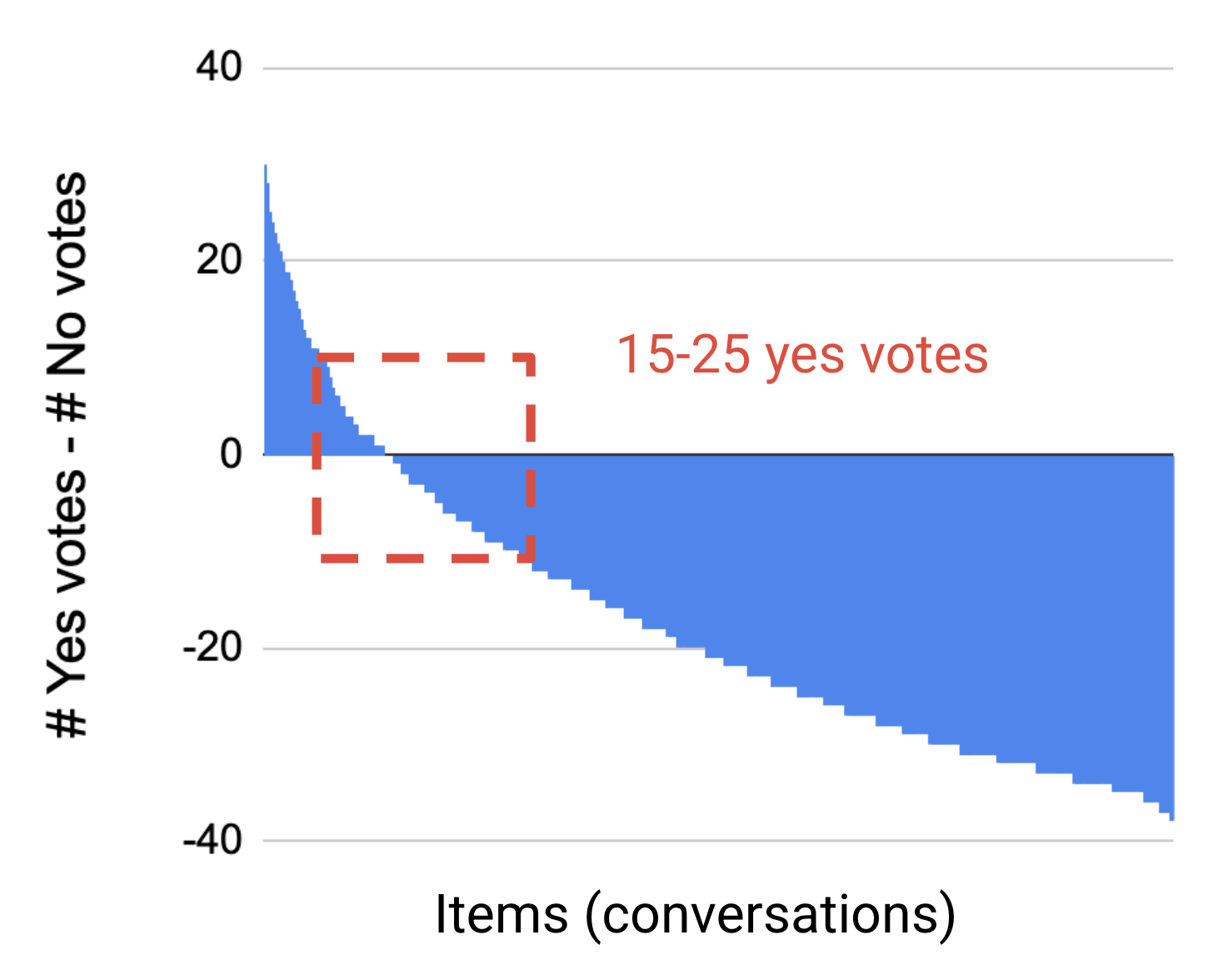}
    \includegraphics[width=0.53\textwidth]{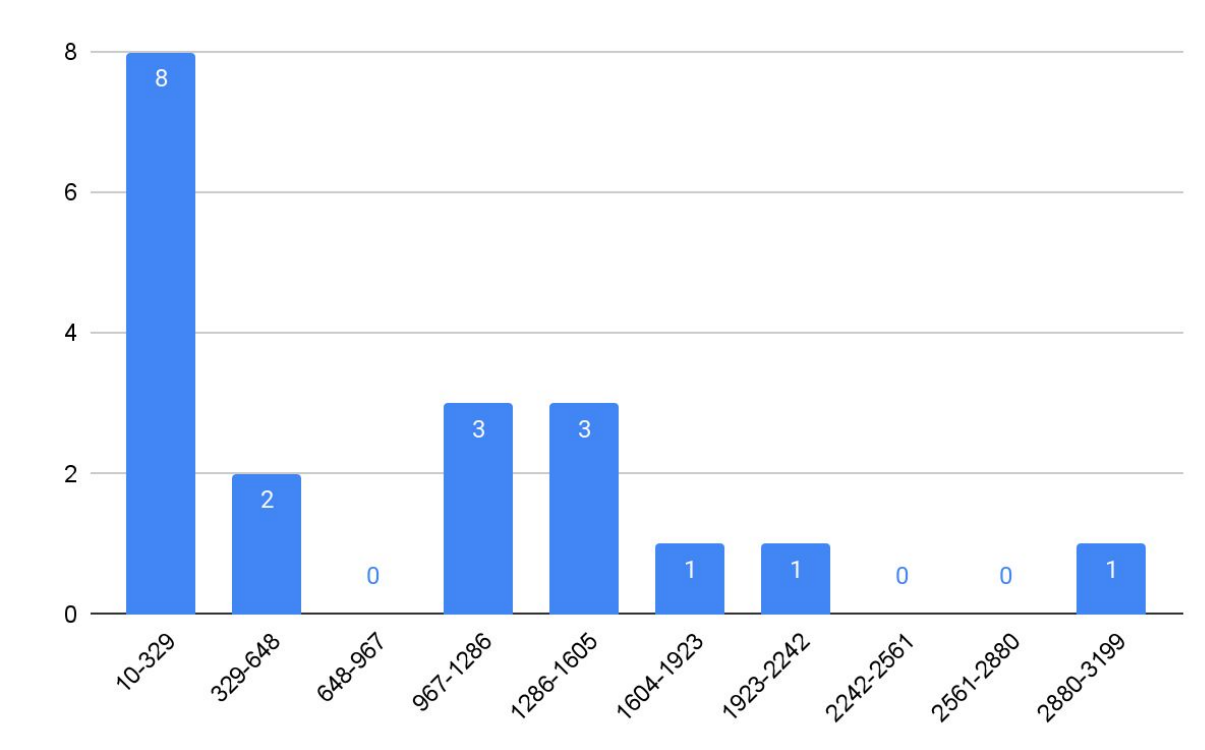}
    \caption{\small \textbf{Left side:} Conversations are arranged horizontally, ranked by the difference between \emph{Unsafe} and \emph{Safe} votes. The y-axis shows this difference. The red square points to roughly a quarter of the conversations with nearly equal numbers of \emph{Unsafe} and \emph{Safe} votes. \textbf{Right side:} Histogram of the number of times each of the 19 raters who rated items in both phases disagreed with themselves.}
    \label{fig:combined}
\end{figure}

\paragraph{Unreliability of gold labels:} The left side of Fig. \ref{fig:combined} shows the difference between the number of raters saying \emph{Unsafe} vs. \emph{Safe} for each of the 990 conversations in our data. For around a quarter of the conversations, the number of \emph{Unsafe} and \emph{Safe} responses per conversation are quite similar, i.e., between 15-25 votes on either side. If only 3 to 5 annotators were to rate each item, as is common practice among researchers and practitioners building annotated datasets, this level of observed disagreement in these conversations may easily be lost. This suggests that majority-based (or even `unanimous') gold labels may be unreliable for a significant portion of the data, if the replication per item is low. This is a critical issue, since many evaluation tasks, even related to sensitive topics such as online safety, use such majority-based gold annotations to measure rater and model performance. 

\paragraph{Inadequate intra-rater consistency:}
We now test whether the 19 raters who were present in both phases were consistent in their ratings. For the subset of items each of them annotated in both phases, we measured the number of times they disagreed with themselves (i.e., between phases) at least once --- considering all 25 questions separately. The right side of Fig. \ref{fig:combined} shows the histogram of disagreements for these 19 raters. Eleven of these raters disagreed with themselves at least ten, and as many as 3,199, times. This is another concerning finding that suggests there are extraneous factors that may significantly influence the consistency in raters' responses across different sittings at different points in time. 

\paragraph{Disparate within-group coherence across subgroups:}
Despite the issues with consistency and reliability, we observed significant patterns in rater behaviour within and across the various subgroups we considered. For this analysis, we modeled each annotator's response to a conversation as a 72-dimensional \textit{response vector} that captures the one-hot encoding of the \{UNSAFE, SAFE, UNSURE\} answers for each of the 24 safety questions (Q2-Q4). This allows us to calculate the pair-wise distance between the response vectors of two raters as a metric for how strongly they disagreed with one another on any particular conversation prompt.

\begin{figure}[h!]
    \centering
    \includegraphics[width=.97\textwidth]{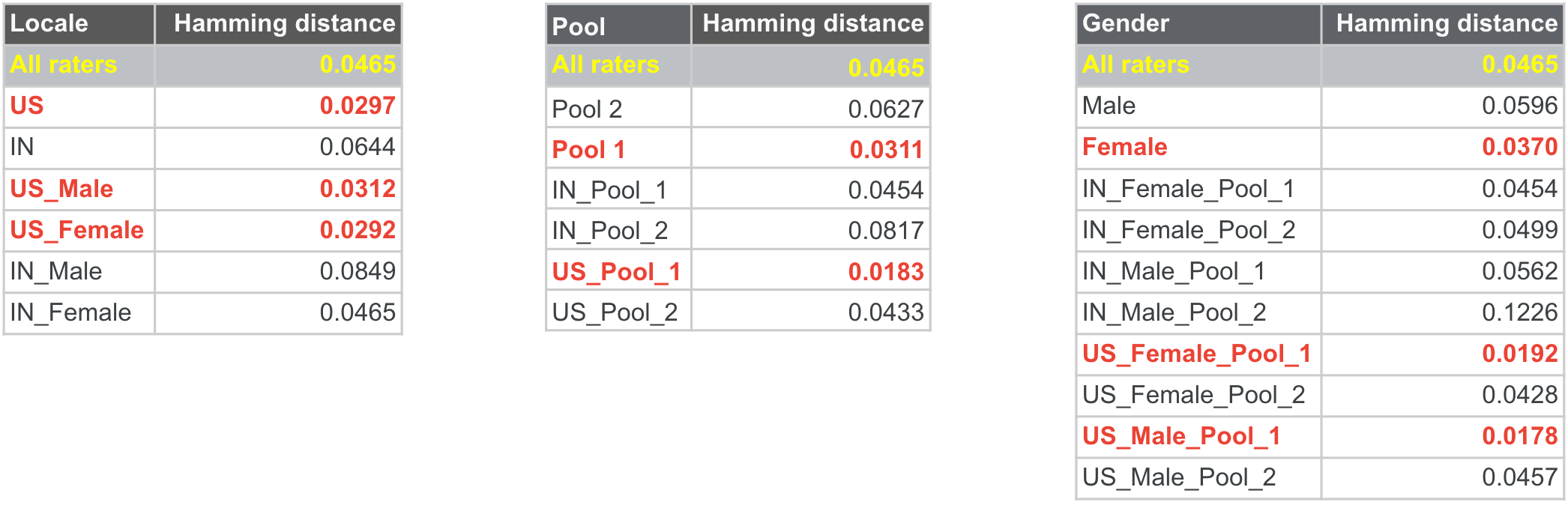}
    \caption{The hamming distance metric on gender, locale and pool slices}
    \label{fig:hamming}
\end{figure}

Fig.~\ref{fig:hamming} shows the average \emph{hamming distance} between all pairs of response vectors for each conversation, averaged across raters within different subgroups of raters. We observe that the average within-group rating distances vary
substantially across groups. 
 
Lower hamming distance between a subgroup and \emph{All raters} means that the subgroup is consistent within itself and different than all raters. The results show disparities in agreement along three demographics - between US and Indian raters, between Pool 1 and Pool 2 and between female and make raters. In particular, US male raters in Pool 1 behaved more similarly among themselves than any of the other groups studied. 

\paragraph{Cross-group differences between subgroups:}
The above analysis provides only a partial picture, one that captures within-group distances but says nothing about whether the rating behaviors of a certain group of raters is more likely to be similar to others in the same subgroup than to those outside the subgroup. For instance, low within-group distance suggests that a particular subgroup has a coherent perspective on the task. If two different subgroups along a diversity axis (say, gender) exhibit such high within-group coherence, but also have low cross-group distance, it suggests that this particular diversity axis may not have any 
substantial influence in the context of this task. However, if two groups that have low within-group distance, also have high cross-group distance, it suggests that the diversity axis is a 
substantial
differentiator for the task. 

\begin{figure}
    \centering
    \begin{subfigure}[t]{0.3\textwidth}
        \vskip 0pt
        \includegraphics[width=\textwidth]{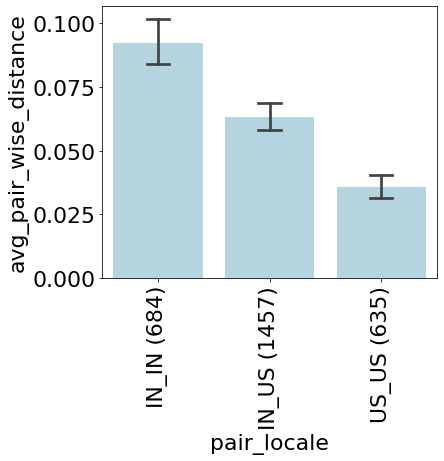}
    \end{subfigure}\hspace{5mm}
    \begin{subfigure}[t]{0.3\textwidth}
        \vskip 0pt
        \includegraphics[width=\textwidth]{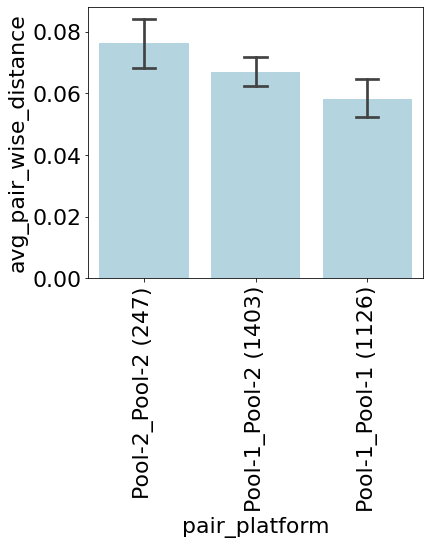}
    \end{subfigure}\hspace{5mm}
    \begin{subfigure}[t]{0.3\textwidth}
        \vskip 0pt
        \includegraphics[width=\textwidth]{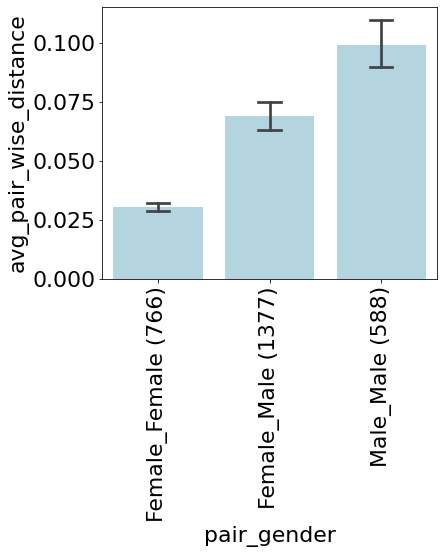}
    \end{subfigure}
    \caption{Average pairwise hamming distance across different locale, platform, gender slices. The number of pairs within each group is specified in parenthesis.}
    \label{fig:cosine}
\end{figure}

Fig.~\ref{fig:cosine} shows the within-group and cross-group distance along the locale (IN vs. US), pool (Pool 1 vs. Pool 2), and gender axes. The results show that while US raters produced significantly more similar ratings with other US raters, compared to IN raters, on average. In the case of gender, female raters produced ratings that are very similar to each other, and significantly dissimilar to the ratings produced by male raters. Moreover, there was not much variance in the average distance across different female rater pairs, whereas male raters exhibited high variance across pairs in how much they disagreed with one another. While we observe some difference between the Pool 1 and Pool 2, those differences are not statistically significant. 


\section{Reflections}
\label{discussion} 

In this paper we are excited to share just a few of the high-level results from the presented work. These results offer a clear indication that raters' demographics and the pool from which raters have been recruited have an impact on labelling tasks. Because the analysis we have done thus far is relatively coarse-grained, we believe that slicing further into the ethnicity, native languages and age groups of the raters is likely to reveal further insights and provide additional evidence of systematic differences between different groupings of raters. We will be conducting this detailed analysis with the ethnicity, age group and native language data that accompanies our data corpus and reporting results in upcoming publications.

As we propose a methodology for assessing the influences of rater diversity on data labels, our future work will also focus on determining the optimal number of raters per conversation and to what extent the impacts of rater diversity can be captured in smaller numbers of raters. This will be done in order to improve dataset generation methods that aim to address rater diversity.

Finally, we recognize more work is needed to help distinguish \emph{good} from \emph{bad} disagreement. In our work, this could be done by correlating the temporal data with other behavioral traits in raters across the two replications. Ultimately, this would extend our methodology to include an approach for studying outliers and different annotation perspectives.

\bibliographystyle{plain}
\bibliography{neurips_2022_hegm}

\end{document}